# Sleep Modulation: The Challenge of Transitioning from Open Loop to Closed Loop

Guisong Liu, Jiansong Zhang, Yinpei Luo, Guoliang Wei, Shuqing Sun, Shiyang Deng, Pengfei Wei*, Nanxi Chen*

*Abstract*—**Sleep disorders have emerged as a critical global health issue, highlighting the urgent need for effective and widely accessible intervention technologies. Non-invasive brain stimulation has garnered attention as it enables direct or indirect modulation of neural activity, thereby promoting sleep enhancement in a safe and unobtrusive manner. This class of approaches is collectively referred to as sleep modulation. To date, the majority of sleep modulation research relies on open-loop paradigms with empirically determined parameters, while achieving individual adaptation and modulation accuracy remains a distant objective. The paradigm-specific constraints inherent to open-loop designs represent a major obstacle to clinical translation and large-scale deployment in home environments. In this paper, we delineate fundamental paradigms of sleep modulation, critically examine the intrinsic limitations of open-loop approaches, and formally conceptualize sleep closed-loop modulation. We further provide a comprehensive synthesis of prior studies involving five commonly employed modulation techniques, evaluating their potential integration within a closed-loop framework. Finally, we identify three primary challenges in constructing an effective sleep closed-loop modulation system: sensor solution selection, monitoring model design, and modulation strategy design, while also proposing potential solutions. Collectively, this work aims to advance the paradigm shift of sleep modulation from open-loop toward closed-loop systems.**

## I. INTRODUCTION

Sleep is a fundamental physiological state that occupies nearly one-third of the human lifespan. Sufficient sleep serves as a critical foundation for essential biological functions, including immune system regulation, maintenance of metabolic homeostasis, and facilitation of tissue repair [1]. However, the six principal categories of sleep disorders, with insomnia being the most prevalent, constitute a major public health concern affecting a wide population and frequently co-occurring with various physical and psychiatric conditions [2]. Epidemiological studies report that, during the COVID-19 pandemic, the global prevalence of insomnia symptoms ranged between 20% and 45%, impacting more than one billion individuals [3]. Insomnia further exhibits a bidirectional association with anxiety and depression [4], thereby exerting an indirect yet substantial influence on population-level mental health. Among older adults, over 60% of those with cognitive impairment or dementia present with one or more forms of sleep disorder [5]. Patients with Mild Cognitive Impairment demonstrate significant alterations in sleep macroarchitecture compared with age-matched healthy controls [6], changes that contribute to the accumulation of amyloid-beta, a defining pathological marker of Alzheimer's disease [7]. Furthermore, chronic sleep deprivation and sustained sleep disturbances are strongly linked to an elevated risk of numerous conditions, including hypertension, diabetes, obesity, depression, cardiovascular disease, and stroke [8]. Collectively, sleep disorders represent an urgent global health burden with far-reaching clinical and societal implications [9].

Insomnia, the most commonly diagnosed sleep disorder in clinical practice [2], has traditionally been managed with sedative-hypnotic agents such as benzodiazepines [10]. However, concerns regarding adverse effects, including tolerance, dependence, and addiction potential, remain a subject of considerable debate [11]. Although orexin receptor antagonists have recently emerged as promising pharmacological alternatives, their long-term safety profile has yet to be fully established [12]. Current clinical guidelines recommend Cognitive Behavioral Therapy for Insomnia as the first-line treatment [13]. Nonetheless, its widespread implementation remains limited by practical barriers, such as a shortage of trained providers and the substantial time

This work was supported by the National Natural Science Foundation of China under Grant L2424238. *(Corresponding author: Pengfei Wei and Nanxi Chen.)*

Guisong Liu and Pengfei Wei are with the State Key Laboratory of Digital Medical Engineering, School of Biological Science and Medical Engineering, Southeast University, Nanjing 210000, China (e-mail: 230258331@seu.edu.cn; 101014012@seu.edu.cn).

Nanxi Chen is with the Department of Biomedical Engineering, Bioengineering College, Chongqing University, Chongqing 400100, China (e-mail: 20181901003g@cqu.edu.cn).

Jiansong Zhang is with the School of Computer Science & Software Engineering, Shenzhen University, Shenzhen 518000, China (e-mail: 2453103003@mails.szu.edu.cn).

Yinpei Luo is with the College of Health Management, Xihua University, Chengdu 610039, China (e-mail: yinpluo@163.com).

Guoliang Wei is with the College of Traditional Chinese Medicine, Chongqing University of Chinese Medicine, Chongqing 402760, China (e-mail: 1305416129@qq.com).

Shuqing Sun is with the Chongqing Xishan Science&Technology Co.,Ltd, Chongqing 404100, China (e-mail: 1525873009@qq.com).

Shiyang Deng is with the School of Basic Medical Sciences, China Medical University, Shenyang 110000, China (e-mail: dengshiyang@cmu.edu.cn).

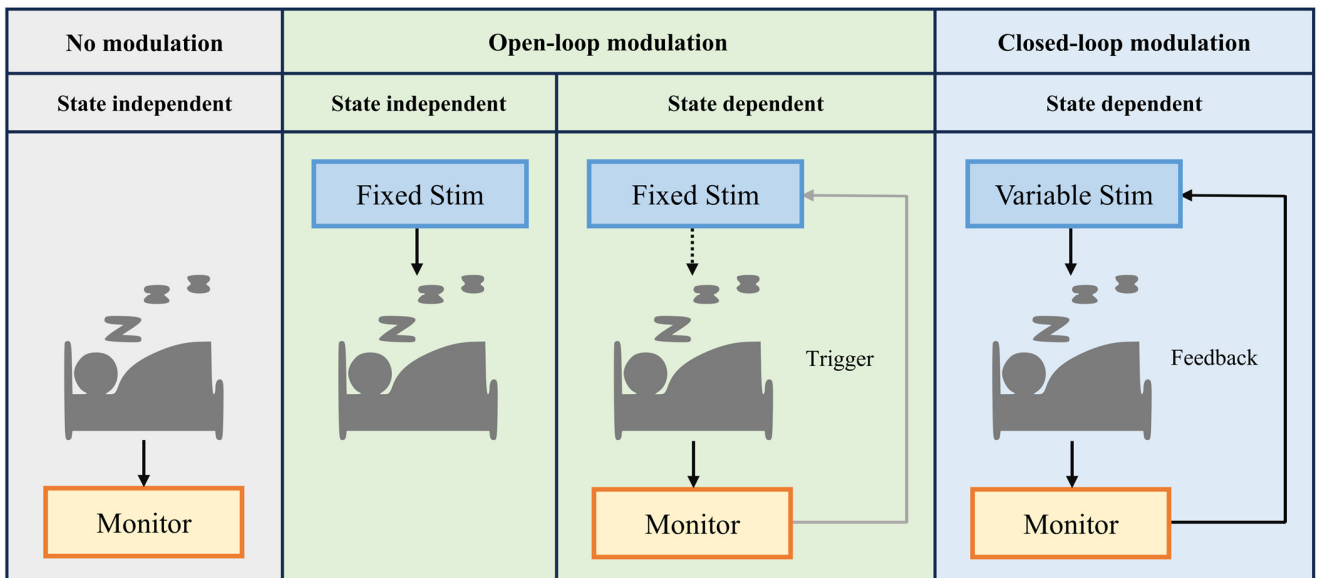

Fig.1.  Four Paradigms of Sleep Monitoring and Modulation.

investment required over the course of therapy [14]. These limitations underscore the pressing need for alternative interventions that are both effective and broadly accessible.

Physical stimulation has gained attention as an adjunctive approach owing to its safety, non-invasiveness, and operational simplicity. By delivering exogenous stimuli to directly or indirectly modulate neuronal activity and thereby influence sleep physiology, this strategy can be broadly conceptualized as *sleep modulation*. Over the past several decades, extensive research has explored various forms of sleep modulation [15], encompassing primarily electrical stimulation, magnetic stimulation, and sensory stimulation. Nevertheless, the findings remain inconsistent, and no consensus has been reached regarding optimal stimulation parameters or timing. Consequently, the therapeutic efficacy of sleep modulation has yet to be conclusively established. We propose that these limitations largely arise from the inherent constraints of the stimulation paradigms employed to date.

Conventional stimulation paradigms generally rely on predefined parameters and are implemented in either a state-dependent or state-independent manner, collectively classified as sleep open-loop modulation. In the following section, we examine in detail the limitations of this approach in the context of sleep modulation. Recently, increasing attention has been directed toward the development of sleep closed-loop modulation systems [16], [17], [18], [19], [20], [21], [22], [23]. The principal advantage of closed-loop designs lies in their ability to continuously acquire real-time physiological signals and adaptively optimize stimulation parameters and timing. This dynamic responsiveness allows for greater precision and flexibility, with the potential to achieve superior therapeutic outcomes. Despite these advantages, several critical challenges remain. First, a rigorous and widely accepted operational definition of sleep closed-loop modulation has yet to be formulated. Second, key conceptual distinctions such as those between open-loop, closed-loop, state-dependent, and adaptive modulation remain ambiguously defined. Third, most current strategies are empirically designed and lack methodological sophistication, resulting in limited reliability and low temporal resolution. Finally, the question of how to optimally select and implement modulation strategies remains unresolved. Addressing these gaps is essential to advancing the transition of sleep modulation from open-loop to closed-loop paradigms.

The contributions of this paper are summarized as follows:
*1)* This study systematically categorizes sleep monitoring and modulation devices into four groups based on morphological configurations and clarifies several frequently conflated concepts. It further establishes the formal definition of sleep closed-loop modulation and outlines key evaluative criteria for this paradigm, thereby providing a conceptual foundation for paradigm selection in the field.
*2)* A comprehensive survey and critical evaluation of prevalent sleep modulation methodologies is conducted, assessing their feasibility for integration into closed-loop systems. This analysis identifies essential operational considerations, delineates effective parameter ranges, and offers methodological guidance for the design of sleep closed-loop modulation systems. In addition, forward-looking discussions are presented to enhance the precision and reliability of modulation strategies.
*3)* Three core challenges in implementing sleep closed-loop modulation systems are addressed, alongside proposed solutions: (i) Sensor strategies: a low-burden, high-accuracy framework based on prefrontal electroencephalography (EEG) augmented by multimodal inputs; (ii) Monitoring models: a four-dimensional design framework emphasizing lightweight architecture, real-time performance, transferability, and interpretability; and (iii) Modulation strategies: a reinforcement learning (RL)-driven conceptual framework. Collectively, these contributions aim to overcome key implementation bottlenecks in sleep closed-loop modulation.

This work aspires to serve as both a reference and a catalyst for researchers, engineers, and clinicians in sleep modulation, with the ultimate goal of facilitating future investigations and advancing the industrial translation of next-generation sleep modulation technologies.

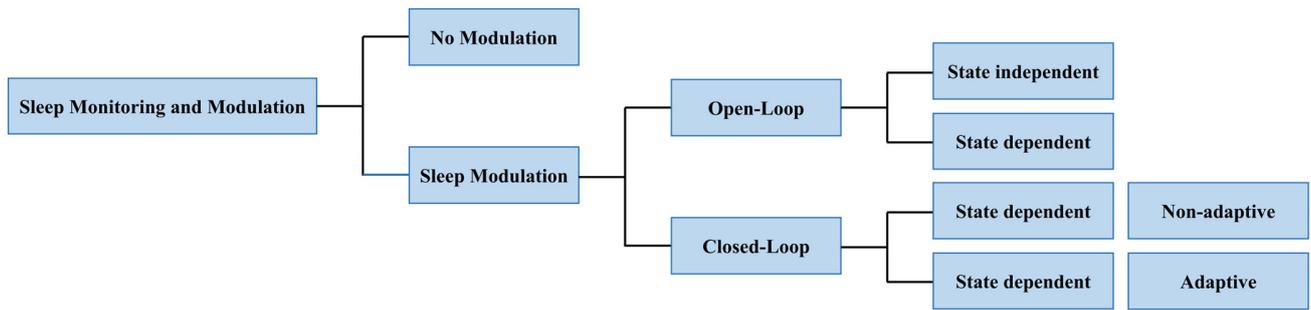

Fig.2. Concepts That Are Easily Confused and Their Corresponding Relationships

## II. DEFINITION OF SLEEP MODULATION

### A. Sleep Monitoring and Sleep Modulation

Polysomnography (PSG) remains the routine practice and gold standard for sleep monitoring, offering multi-channel acquisition for the accurate diagnosis and evaluation of diverse sleep disorders. Among its core functions, sleep staging is particularly critical. According to the guidelines of the American Academy of Sleep Medicine, each 30-second epoch is classified into one of five stages: wake (W), non-rapid eye movement (N1, N2, N3), or rapid eye movement (REM) sleep [24]. This process is typically performed manually by trained physicians, who annotate epochs based on full-night electrophysiological signals recorded by PSG, including EEG, electrooculography (EOG), and electrocardiography (ECG), among others.

Despite its widespread clinical use, PSG is limited by factors such as complex electrode placement, high equipment costs, and the potential to interfere with natural sleep, thereby restricting its applicability in home-based environments. To address these challenges, wearable sleep monitoring devices have been developed by simplifying conventional PSG systems to achieve portable, low-burden assessments outside clinical settings [25]. These devices not only facilitate personal health management and early detection of sleep disorders but also support advances in sleep modulation.

With the growing diversity of wearable systems and experimental paradigms, the field of sleep monitoring and modulation is becoming increasingly heterogeneous. Despite substantial technological progress, systematic efforts to establish unified standards and taxonomies remain insufficient. This lack of harmonization has led to inconsistencies in methodology and evaluation across sleep modulation studies, posing significant challenges for device design, performance benchmarking, and experimental reproducibility. To address this gap, the present study proposes a classification framework that categorizes existing sleep monitoring and modulation devices into four representative configuration types, as illustrated in Fig. 1.

#### 1) Category 1
Sleep monitoring exclusively, without any stimulation.

#### 2) Category 2
Fixed stimulation delivery exclusively, without sleep monitoring. Stimulation triggers are predetermined and operate independently of physiological state.

#### 3) Category 3
Predefined and fixed Stimulation triggered by sleep monitoring data. Post-activation, the stimulation strategy remains static regardless of subsequent monitoring data.

#### 4) Category 4
Stimulation is triggered and adjusted in real-time based on sleep monitoring feedback (fixed or dynamic stimulation), which constitutes the sleep closed-loop modulation paradigm proposed in this work.

Based on our definition above, open-loop systems can be classified into two categories: state-independent systems that do not require sleep monitoring, and state-dependent systems that are triggered based on such monitoring. Herein, state-dependent denotes whether stimulation delivery relies on real-time sleep state monitoring data. It should be emphasized that state-dependent and closed-loop operation are not entirely equivalent. While all closed-loop systems are inherently state-dependent, not all state-dependent systems operate in a closed-loop manner.

Additionally, the adaptive concept referenced in the literature is typically confined to closed-loop systems. Closed-loop systems can be further categorized into non-adaptive and adaptive variants, contingent on whether modulation strategies possess the capability to recognize sleep physiological and environmental variations, and adjust inherent modulation protocols accordingly. This taxonomy not only establishes fundamental paradigms in sleep modulation but also mirrors the historical development of sleep monitoring and modulation technologies. The relationships among these often-confused concepts are illustrated in Fig. 2.

### B. Sleep open-loop modulation

Conventional sleep modulation typically employs an Open-loop Paradigm [26], [27], [28], [29], [30], [31], [32]. In this approach, stimulation parameters (e.g., intensity, frequency, duration) are preset according to predetermined protocols, often derived from empirical heuristics or prior literature, and remain constant across individuals and throughout the session duration. Such a priori parameters generally represent suboptimal subsets of the full parameter space. A representative implementation involves continuous all-night administration of constant-intensity pink noise stimulation [33], disregarding dynamic changes in the subject's physiological state.

However, neuromodulation efficacy exhibits significant dependence on individual variability, brain state, and task context [34], [35]. The open-loop paradigm treats sleep as a static, time-invariant system, exhibiting inherent limitations. The primary constraints include (Fig.3):

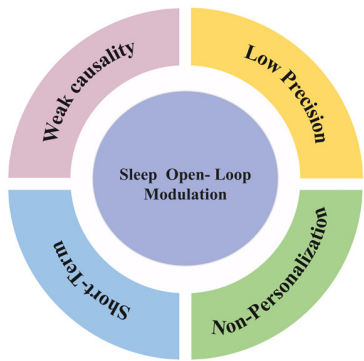

Fig.3.   Defects of Sleep Open-Loop Modulation.

### 1) Weak Causality

Open-loop paradigms demonstrate insufficient causal linkage between modulation and sleep improvement. While comparative studies evaluate sleep enhancement differences between stimulated and control groups, confounding factors remain poorly controlled across studies employing identical protocols [36]. These compromises result in low levels of predictability and reproducibility, undermining causal inference regarding modulation efficacy.

### 2) Low Precision

The inability to dynamically adapt modulation parameters to real-time physiological states fundamentally constrains the precision and granularity of all-night sleep modulation. Predefined parameters may prove ineffective or potentially detrimental under specific conditions.

### 3) Short-Term Focus

By delivering static stimulation without acquiring physiological feedback, open-loop approaches cannot capture longitudinal biomarker evolution during modulation cycles. This precludes assessment of cumulative effects and investigation of long-term neuroplasticity [37], resulting in inherently short-term outcome evaluation.

### 4) Non-Personalization

Mechanically applied identical parameters fail to accommodate pervasive inter- and intra-individual variability [38]. Consequently, they exhibit limited generalizability across individuals or even within the same individual across different periods, necessitating personalized parameter optimization.

These limitations intrinsically constrain the efficacy ceiling of sleep open-loop modulation. Conceptually analogous to a write-only brain-computer interface (BCI), this unidirectional paradigm contradicts the prevailing trend toward bidirectional BCIs [39], [40], [41]. With advancements in non-invasive neuromodulation, transitioning to closed-loop paradigms represents a critical step in advancing research on sleep modulation [37], [42], [43], [44].

### C. Sleep closed-loop modulation

The concept of closed-loop originates in control theory, where its distinction from open-loop systems lies in the utilization of feedback mechanisms that enable system outputs to iteratively inform and adjust inputs. Formally, a closed-loop control process can be expressed as:

$$u(t) = K[r(t) - y(t)] \qquad (1)$$

Where $r(t)$ denotes the reference signal, representing the desired output of the system, $y(t)$ is the measured system output, and $u(t)$ represents the control input. $K[\cdot]$ denotes controller function. The feedback term $r(t) - y(t)$ continuously drives the system toward the desired state, enabling adjustment based on real-time performance.

Within neuroscience, the closed-loop paradigm establishes dynamic interactions between neural circuits and the external environment or devices through real-time feedback [45]. This paradigm provides a robust framework for causal inference [23], [39] and has been widely adopted in neuroengineering applications [46], [47], [48], [49], [50], [51], where it has consistently improved both user experience and intervention efficacy. Extending this framework to sleep research holds considerable promise for advancing sleep modulation to an unprecedented level [52]. A recent review of 148 studies investigating "sleep and stimulation" identified only 20 that employed closed-loop methodologies [53].

To date, no universally accepted formal definition of sleep closed-loop modulation has been established [53], [54], [55]. Drawing upon a comprehensive review of prior work and consultation with academic and industry experts, we propose the following definition: *Sleep closed-loop modulation refers to a system that continuously acquires physiological information with minimal disruption to natural sleep, performs real-time analysis of relevant biomarkers, and delivers physical stimulation in a feedback-controlled manner. The system modifies stimulation strategies in response to ongoing physiological dynamics, thereby enhancing sleep quality or other associated physiological functions.*

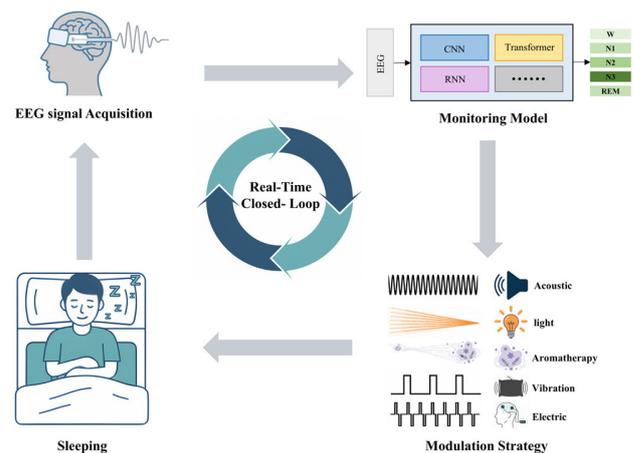

Fig.4.   Schematic representation of a typical sleep closed-loop modulation system. The framework comprises three core components: signal acquisition, monitoring model, and modulation strategy. During sleep, physiological signals are continuously captured through wearable hardware. The monitoring model processes these signals in real time to evaluate relevant biomarkers, which in turn inform the modulation strategy for dynamically selecting and adjusting stimulation modalities and parameters. These adjustments subsequently alter physiological states, thereby completing the feedback cycle that characterizes a closed-loop system.

Fig.4 illustrates an example of a typical sleep closed-loop modulation system.To ensure conceptual clarity, this study further delineates three essential verification criteria for authentic closed-loop operation in the context of sleep modulation:

1)  *Real-time monitoring of physiological parameters during sleep;*
2)  *Demonstrable modulation impact on sleep states or biomarkers;*
3)  *Continuous optimization of modulation strategy based on feedback (the critical differentiator).*

Consequently, state-dependent triggering without feedback mechanisms does not constitute closed-loop modulation [17]. Moreover, the objectives of sleep modulation differ substantially across clinical populations. For individuals with insomnia, the primary aims are to reduce sleep onset latency and minimize nocturnal awakenings [56]; In contrast, for cognitively impaired older adults, the principal target is to enhance slow-wave sleep in order to support declarative memory consolidation [57]. Thus, modulation targets extend beyond mere "sleep macrostructure" adjustment. In a word, the objective of sleep modulation should be actively restoring or enhancing the cognitive and physiological benefits typically acquired through natural sleep. Crucially, all outcomes must demonstrate beneficial effects. Temporally, our definition restricts modulation to sleep periods (including pre-sleep in-bed intervals), establishing real-time closed-loop operation. No empirical evidence currently exists for closed-loop in stages implementations, making real-time operation essential for conceptual clarity. Biomarkers must satisfy two conditions: (i) Correspondence to modulable behavioral or functional capacities; (ii) Precise responsiveness to intervention. Valid biomarkers encompass not only sleep staging, but also slow-wave oscillations [58], spindle activity [59], [60], [61], and heart rate [62]. Different systems may leverage specific biomarkers to guide the application or refinement of modulation strategies [19], [59], [62], [63], [64].

We further derive four fundamental design tenets for sleep closed-loop modulation systems: Safety: non-negotiable prerequisite (e.g., rigorous assessment of cumulative electrical stimulation duration and intensity thresholds). Scientific Validity: mechanistic transparency of modulation methodology. Efficacy: demonstrated effectiveness through controlled clinical validation. Comfort: absence of physical or psychological discomfort during operation.

## III. METHODS FOR SLEEP CLOSED-LOOP MODULATION

Exogenous stimulation constitutes a fundamental methodological framework for probing sleep homeostasis by introducing targeted perturbations. Although prior studies have reviewed diverse modulation strategies from multiple perspectives [15], [52], [55], [56], [65], [66], [67], the present section systematically evaluates their relevance and feasibility within closed-loop implementations. Specifically, we provide a concise review of five modulation modalities with practical applicability, emphasizing their potential for integration into closed-loop systems. For clarity, key features including experimental paradigms, validated parameter ranges, participant characteristics, and reported intervention outcomes are systematically tabulated. Subsequently, the feasibility associated with integrating various modulation approaches into closed-loop systems are discussed, along with a critical analysis of methodological limitations and the essential considerations for implementation. Finally, the section concludes with a forward-looking discussion that outlines proposals to advance methodological research in sleep modulation.

### A.  Acoustic stimulation

#### 1)  Empirical evidence for acoustic modulation of sleep

The gentle pitter-patter of rain often induces drowsiness [68], the rhythmic lapping of waves promotes deep relaxation [69], and even the hum of a fan can cause sleepiness. These phenomena illustrate the subtle influence of sound on sleep modulation, as specific acoustic frequencies can modulate sleep-wake centers in the brain via the auditory pathway. Reported effective auditory stimuli for sleep facilitation include white noise, pink noise, and binaural beats.

White noise, defined by its uniform energy distribution across the frequency spectrum, has been shown to exert notable sleep-promoting effects. In infants, exposure to 93 dB white noise was associated with a threefold increase in the likelihood of sleep onset [70], while continuous presentation of 75 dB white noise throughout the night reduced resistance to sleep initiation and decreased nocturnal awakenings [71]. Similarly, white noise composed exclusively of intrauterine sounds increased total sleep duration and improved sleep efficiency in preterm infants [72]. In clinical populations, administration of 40-50 dB white noise for one hour significantly reduced Pittsburgh Sleep Quality Index (PSQI) scores in patients with coronary heart disease admitted to a Coronary Care Unit [73]. Collectively, these examples demonstrate white noise's efficacy in calming individuals for sleep onset and improving sleep quality. Its broadband properties mask disruptive external sounds, thereby increasing arousal thresholds [74].

Pink noise exhibits stronger low-frequency components and is the most prevalent noise type in natural environments. Zhou et al. demonstrated that steady pink noise improved individual sleep quality by reducing EEG complexity and increasing stable sleep duration [33], while Kawada et al. observed a mean reduction of 9.5 minutes in sleep onset latency among four participants exposed to continuous 60 dB pink noise relative to baseline [75]. One hour of pre-sleep pink noise exposure significantly decreased PSQI scores in healthy university students [69]. Research suggests pink noise increases EEG alpha and low-beta wave activity, frequency bands associated with relaxation and comfort [76].

Other auditory stimulation paradigms have also been investigated. Binaural beats, for example, have demonstrated potential for modulating sleep architecture: Nantawachara et al. applied 3 Hz binaural beats during N2 sleep, which accelerated the transition to N3 and prolonged its duration [77]. Likewise, Fan et al. reported that continuous 0.25 Hz binaural beat stimulation shortened latencies to N2 and N3 during naps, though the underlying mechanisms remain insufficiently

TABLE I ACOUSTIC STIMULATION RESEARCH COLLATION.

| Method | Study design | Stimulus parameters | Subject | Result |
|---|---|---|---|---|
| Acoustic | Randomized controlled | white noise, 72.5 dB for the first 30s ,67 dB for the remaining 4 min | 40 healthy neonates (aged 2-7 day) | 5 min nap, trial group: 80% fell asleep control group: 25% fell asleep [70] |
| Acoustic | Pre–post contrast | white noise, 75 dB | 4 toddlers, Easy to wake up at night (aged 13-23 month) | Whole night and nap, 3 toddlers sleep better [71] |
| Acoustic | Single-blind control | white noise, 40-50 dB, duration 1h | 60 patients admitted to the coronary care in CCU, 33 male, 27 female | Whole night, PSQI(-) [73] |
| Acoustic | randomized controlled | white noise, ≤50 dB,  duration 15 min | 120 premature infants in NICU, (aged 28-37 weeks) | Whole night, sleep duration(+), sleep efficiency(+), awakening(-), WASO(-) [72] |
| Acoustic | Pre–post contrast | pink noise, 20-40 dB | 40 healthy adults, 22 male, 18 female (mean age 23.4 ±2.9) | Whole night, stable sleep time(+) [33] |
| Acoustic | Pre–post contrast | pink noise, 60 dB | 4 students, 3 male, 1 female (aged 19-21) | Whole night, sleep latency(-) [75] |
| Acoustic | Pre–post contrast | pink noise, 1h before sleep | 120 healthy students, 58 male, 62 female (mean age 22.9±1.63) | Whole night, PSQI(-) [69] |
| Acoustic | Control | Binaural Beat, 3 Hz, 60 dB, only when the first N2 until N3 appears | 24 healthy adults, 13 male, 11 female (mean age 24.1 ±2.54) | Whole night, N2(-), N3(+), N3 latency(-) [77] |
| Acoustic | Pre–post contrast | Binaural Beat, 0.25 Hz | 12 healthy adults, 6 male, 6 female (mean age 25.4 ±2.6) | 90 min nap, N2 and N3 latency(-) [78] |

CCU: Coronary Care Unit; NICU: Neonatal Intensive Care Unit; PSQI: Pittsburgh Sleep Quality Index; WASO: Wake After Sleep Onset.

clarified [78]. Furthermore, closed-loop auditory stimulation has emerged as a promising technique, wherein slow-wave oscillations are detected in real time and stimuli are delivered at specific phases (e.g., the up-state) and frequencies (e.g., <1 Hz), thereby enhancing slow-wave activity and consolidating declarative memory [79]. In addition, auditory cues have been applied during REM sleep for targeted memory reactivation, with demonstrated efficacy in augmenting imagery rehearsal therapy for nightmares [80].

*2) The feasibility of closed-loop integration*

Auditory stimulation presents several practical advantages, including low implementation complexity, seamless integration through headphones, and high user acceptability. Its capacity for precise temporal control and targeted delivery [81] has supported its early adoption in closed-loop paradigms [16]. Nevertheless, auditory input is inherently double-edged. Inappropriate environmental noise can disrupt sleep architecture and trigger arousals [82], with particularly pronounced effects in vulnerable populations such as older adults, children, and shift workers [83]. The World Health Organization accordingly recommends maintaining average nighttime noise levels below 40 dB [84], as excessive intensity not only disturbs sleep but also poses risks of hearing damage. Although emerging studies indicate that sound-based interventions may alleviate insomnia symptoms [85], empirical support for their efficacy in promoting general sleep enhancement remains limited. Two recent systematic reviews concluded that the overall quality of evidence for auditory stimulation as a sleep-improving modality is low [86], [87]. Consequently, a more rigorous evaluation of auditory stimulation during sleep is warranted. Table I summarizes the acoustic stimulation research we have compiled.

*B. light therapy*

*1) Empirical evidence for light modulation of sleep*

At sunrise, we labor, at sunset we repose, reflecting the essential role of light as a primary zeitgeber in the regulation of circadian rhythms. Fundamental investigations have demonstrated that light signals are transduced via intrinsically photosensitive retinal ganglion cells (ipRGCs) and subsequently projected to the suprachiasmatic nucleus (SCN) of the hypothalamus [88]. The SCN synchronizes circadian rhythms with environmental light cycles by modulating melatonin secretion from the pineal gland [89]. This non-visual phototransduction pathway provides the theoretical foundation for light-based interventions in sleep modulation [90]. Studies have shown that ipRGCs exhibit maximal sensitivity to blue light [91], [92]; accordingly, filtering excessive blue wavelengths enhances melatonin secretion and promotes sleep [93], while intense blue-enriched light exposure exerts opposing effects.

Research has demonstrated that morning bright light exposure induces a phase advance in patients with Delayed Sleep Phase Syndrome [94], suggesting its therapeutic potential for correcting circadian misalignment in individuals with insomnia or shift work schedules [95]. In elderly patients with insomnia, daily 45-minute sessions of morning light exposure over two months produced sustained improvements in sleep onset latency and total sleep duration during follow-up [96]. Similarly, two weeks of daily 2-hour bright light exposure improved sleep efficiency and significantly reduced nocturnal wake time in older adults with dementia-related sleep disturbances [97]. Post-treatment phase delays in the 24-hour plasma melatonin concentration profile [98] further provide

TABLE II LIGHT THERAPY RESEARCH COLLATION.

| Method | Study design | Stimulus parameters | Subject | Result |
|---|---|---|---|---|
| Light | Control | 2,500-lx full- spectrum light treatment, 2 h,  6:00- 9:00 a.m; dark goggles in the evening, for 1 week | 36 DSPS patient | Whole night, circadian rhythms of core body temperature and multiple sleep latencies (phase advanced) [94] |
| Light | Control | 10,000 lx light, morning, 45min, for 60 days | 30 insomnias in elderly, 10 male, 20 female, (mean age 64.8± 7) | Whole night, Sleep latency(-), total sleep time (+) [96] |
| Light | Pre–post contrast | 6000-8000 lx bright light, 2h, 08:00-11:00 a.m, for 2 weeks | 11 demented elderly with sleep disturbances, 1 male, 10 female, (mean age 86.1± 8.9) | Whole night, sleep efficiency(+), waking time(-) [97] |
| Light | Control | 2,500- 3,000 lx, full spectrum white light 2 h, for 2 weeks | 14 night-shift men, (aged 21-35) | Whole night, Sleep latency(-), sleep duration(+) [98] |
| Light | Pre–post contrast | artificial dawn, 30 mins before awakening | 16 healthy subjects with difficulty awakening, 8 male, 8 female, (mean age 22.8± 4.6) | Whole night, After waking , subjective sleepiness(-), subjective activation(+) [99] |
| Light | Pre–post contrast | artificial dawn, 30 mins before awakening, 0.001 lux gradually increases to 300 lux | 8 healthy subjects with difficulty awakening, 4 male, 4 female, (mean age 24± 9) | Whole night, After waking, Subjective perceived sleep quality(+), Subjective alertness (+), cognitive performance (+), physical performance(+) [100] |
| Light | Crossover, within-subjects, counterbalanced | red light mask(688 lx) worn during sleep | 30 healthy subjects, 12 males, 18 females, (mean age 30.4± 13.7) | 90 mins sleep, sleep inertia(-) [101] |

DSPS: Delayed Sleep-Phase Syndrome.

direct evidence of light therapy's regulatory effects on circadian melatonin rhythms.

Light therapy has also been shown to mitigate sleep inertia. Artificial dawn simulation administered 30 minutes prior to awakening reduced subjective sleepiness upon waking [99], while another study reported significant enhancements in both physical and cognitive functioning [100]. Comparable effects have been observed with saturated red light exposure under closed-eye conditions, which reduced sleep inertia [101]. Emerging findings further suggest that light exposure may exert effects beyond the visual system: whole-body red light irradiation improved both sleep and endurance performance in elite athletes [102], and cervical near-infrared light collars elicited self-reported improvements in sleep quality and mood [103]. More recently, 40 Hz flickering light stimulation has demonstrated potential for enhancing sleep [104], thereby expanding the repertoire of photobiomodulation approaches.

2) The feasibility of closed-loop integration

A systematic review has demonstrated that light therapy exerts its strongest effects in the management of circadian rhythm disorders and insomnia [105], indicating its potential utility primarily as an adjunctive approach complementing other sleep interventions. Existing protocols, which typically employ fixed intensity and fixed timing, exhibit limited adaptability to interindividual circadian variability and dynamic changes in sleep states. Incorporating closed-loop control strategies may substantially improve therapeutic precision. At the same time, safety considerations remain critical, as prolonged exposure to high-intensity light carries the risk of irreversible ocular damage [106], [107]. Accordingly, future closed-loop investigations should explicitly report comprehensive assessments of ocular safety. Table II summarizes the light therapy research we have compiled.

C.  Aromatherapy

1) Empirical evidence for aromatherapy modulation of sleep

Floral scents are often associated with relaxation and tranquility, whereas refreshing aromas like peppermint can promote alertness and wakefulness. These observations suggest that olfactory stimuli may exert measurable effects on sleep modulation [108]. Aromatherapy utilizes essential oils extracted from plants to improve physical and mental well-being [109], with a history spanning millennia [110]. Administration methods include inhalation, massage, and baths; this discussion focuses solely on inhalation. This modality demonstrates relatively superior efficacy in facilitating sleep [111] and offers practical simplicity. During inhalation, aromatic molecules stimulate olfactory receptors, generating neural impulses that modulate the limbic system [112].

Lavender essential oil is a common choice in aromatherapy due to its low toxicity and allergenicity [113]. As a mild sedative, it enhances deep sleep in healthy individuals [114]. A 12-week regimen of lavender aromatherapy improved sleep quality in middle-aged women with insomnia [115]. A randomized controlled trial demonstrated that inhaling 2% lavender oil for 20 minutes before bedtime improved sleep quality and reduced anxiety levels in coronary patients within an Intensive Care Unit setting [113]. Studies indicate lavender oil increases delta wave activity, indicative of deep sleep [116]. Its active constituents possess sedative and anesthetic properties and may stimulate the parasympathetic nervous system, inducing relaxation [117]. Beyond lavender, other aromatic substances reported to improve sleep quality include valerian [118], damask rose [119], lemongrass and lemon [120], and

TABLE III AROMATHERAPY STIMULATION RESEARCH COLLATION.

| Method | Study design | Stimulus parameters | Subject | Result |
|---|---|---|---|---|
| Aromatherapy | Pre–post contrast | Lavender essential oil, Intermittent inhalation before sleep | 31 healthy adults, 16 male, 15 female, (mean age 20.5±2.4) | Whole night, For all: N3(+); For female: N2(+), REM(-), WASO(-); For male: N2(-), REM(+), WASO(+) [114] |
| Aromatherapy | Prospective and randomized controlled | Lavender essential oil, duration 20 min each time, twice per week, for 12 weeks | 67 midlife females with Insomnia, (aged 45-55) | Whole night, PSQI(-) [115] |
| Aromatherapy | Randomized controlled | 2% Lavender essential oil, duration 20 min before sleep, for 15 days | 60 patients in coronary ICU, 40 male, 20 female, (mean age 53.3±12.1) | Whole night, PSQI(-) [113] |
| Aromatherapy | Randomized controlled | 10% Damask rose essential oil, duration 5-10 min before sleep, for 10 days | 75 mothers with premature newborns in NICU, PSQI ≥5, (mean age 30.2 ± 5.6) | Whole night, PSQI(-) [119] |
| Aromatherapy | Pre–post contrast | lemongrass and lemon blend essential oils, for 7 days | 48 pregnant females | Whole night, PSQI(-) [120] |
| Aromatherapy | Randomized, controlled | essential oils of lavender and peppermint, duration 20 min before sleep, for 7 days | 120 cancer patients, 52 male, 68 female, PSQI ≥5 (mean age 49.4 ±14.5) | Whole night, PSQI(-) [121] |

ICU: Intensive Care Unit; NICU: Neonatal Intensive Care Unit; PSQI: Pittsburgh Sleep Quality Index; WASO: Wake After Sleep Onset.

peppermint [121].

*2) The feasibility of closed-loop integration*

Notably, pronounced sex-related differences have been documented in human olfactory function [122], and multiple studies have reported gender-specific effects of aromatherapy. For example, lavender oil has been shown to prolong N2 sleep while reducing REM sleep and wake after sleep onset in females, whereas opposite effects were observed in males [114]. Similarly, peppermint oil was found to extend REM sleep in females, with no corresponding effect in male participants [123]. Beyond biological differences, evidence suggests that an individual's subjective perception and expectation of an aroma may exert a stronger influence on sleep outcomes than the odorant itself [123], [124], [125]. Methodological limitations also persist, as most studies emphasize the ratio of essential oil to water while neglecting monitoring of ambient essential oil concentration, a parameter directly determining the inhaled dose. The incorporation of closed-loop systems could facilitate real-time monitoring of inhaled concentrations, thereby reducing interindividual variability and subjective bias [126]. Table III summarizes the aromatherapy stimulation research we have compiled.

*D. Vibration*

*1) Empirical evidence for vibration modulation of sleep*

Newborns fall asleep more readily in maternal arms or mechanical cradles [127], and people often experience drowsiness during train travel [128]. Research demonstrates that low-frequency vibration (e.g., train environments [129], cradle rocking [130], linear swaying [131] ) significantly increases sleep propensity and prolongs deep sleep duration.

Drawing inspiration from naturally occurring phenomena, researchers have sought to replicate vibrational patterns to recreate sleep-promoting environments. Kimura et al.

developed a novel mechanical bed and demonstrated that 0.5 Hz vibration shortened sleep onset latency, with imperceptible stimulation producing greater reductions than perceptible vibration, thereby indicating that optimized parameters can effectively facilitate sleep initiation [132]. Similarly, Bayer et al. reported that continuous rocking at 0.25 Hz during 45-minute naps reduced sleep onset latency, decreased N1 duration, and increased N2 duration. Rocking also induced sustained enhancements in slow-wave and spindle activity [133], electrophysiological signatures indicative of deeper sleep. These findings suggest that vibrational stimulation may promote sleep depth by entraining endogenous oscillatory activity. In line with this, Perrault et al. demonstrated that all-night lateral rocking at 0.25 Hz reduced light NREM stages (N1, N2) and prolonged N3 sleep. Furthermore, the temporal clustering of spindles and slow oscillations relative to the rocking cycle provided evidence supporting the hypothesis that rocking entrains sleep oscillations [26]. While most prior studies employed single-frequency stimulation, Himes et al. introduced a beat-frequency vibration bed, in which a 0.5 Hz modulated wave was generated by superimposing two sinusoidal frequencies. This approach reduced sleep onset latency in participants with mild-to-moderate insomnia symptoms [134].

Although vibrational stimulation appears to facilitate sleep through the entrainment of endogenous oscillations, its neural mechanisms remain insufficiently understood. Clinical evidence indicates that vestibular dysfunction can precipitate sleep disturbances [135]. Given the vestibular system's role in mediating balance and motion perception, a prevailing hypothesis posits that vibrational stimulation influences sleep via vestibular pathways, with animal studies implicating the otolith organs as potential mediators [136].

TABLE IV VIBRATION STIMULATION RESEARCH COLLATION.

| Method | Study design | Stimulus parameters | Subject | Result |
|--------|--------------|---------------------|---------|--------|
| Vibration | Pre–post contrast | 0.25 Hz, 10.5 cm lateral excursion | 10 healthy males (mean age 30.1) | 45 min nap, N1(-), N2(+) [133] |
| Vibration | Pre–post contrast | 0.25 Hz, 10.5 cm lateral excursion | 18 healthy adults, 8 male, 10 female (mean age 23.39±1.61) | Whole night, N1+N2(-), N3(+) [26] |
| Vibration | Pre–post contrast | 0.5 Hz, 5 mm amplitude, horizontal or vertical | 10 healthy males (aged 21 -27) | sleep latency(-) [132] |
| Vibration | Pre–post contrast | 0.5 Hz(beat frequency Vibration), a peak acceleration of 0.2 m/s² | 14 mild-moderate insomnia students, 8 female, 6 male, (mean age 22.2±3.0) | sleep latency(-) [134] |
| Vibration | Pre–post contrast | 0.25 Hz, 10 cm amplitude, the acceleration of 0.3 m/s² | 15 healthy males (mean age 25.4±2.99) | 3h nap, N3(+) [130] |

### 2) The feasibility of closed-loop integration

Crucially, not all vibration parameters are conducive to sleep induction. Vibrations at higher frequencies, particularly those exceeding 10 Hz, have been shown to disrupt sleep [129]. Simulated freight train vibrations have demonstrated adverse effects on nocturnal sleep quality [137], while excessive amplitude is associated with discomfort [138], and high acceleration may elicit motion sickness rather than promoting sleep [132]. These findings underscore the necessity of imposing strict safety constraints on vibration stimulation parameters. Moreover, the integration of sleep monitoring to enable real-time adjustment of stimulation parameters, with the aim of entraining stage-specific oscillatory dynamics, offers considerable potential for more precise and adaptive modulation of sleep depth [26]. Table IV summarizes the vibration stimulation research we have compiled.

### E. Transcranial Electrical Stimulation

#### 1) Empirical evidence for transcranial electrical stimulation modulation of sleep

Subtle electrical currents, akin to night dew permeating soil, can nurture the bloom of deep sleep. Transcranial Electrical Stimulation (tES) modulates cortical excitability by delivering weak electrical currents (–2 to 2 mA) to targeted brain regions [139]. By directly influencing cortical activity, tES enables non-invasive modulation of sleep through top-down cortical-thalamic circuits that modulate arousal and sleep [140].

Evidence suggests that tES can enhance memory consolidation during sleep. A landmark study in 2004 demonstrated that intermittent bilateral anodal transcranial direct current stimulation (tDCS) during slow-wave sleep (SWS) in healthy young adults increased sleep depth and facilitated the sleep-dependent consolidation of declarative memory. Anodal tDCS also enhanced EEG slow-wave oscillations (SWO) [58], which play a causal role in memory consolidation processes [141]. A study published in Nature similarly reported that applying slow-oscillatory tDCS (so-tDCS) during SWS enhanced the consolidation of hippocampus-dependent declarative memory in healthy individuals. The oscillation frequency was set to 0.75 Hz to optimally elicit slow-wave oscillations with peak frequencies in the 0.7-0.8 Hz range [141]. These conclusions are supported by meta-analytical evidence [142].

Anodal tDCS has also been shown to sustain wakefulness and improve cognitive performance. In healthy individuals, bifrontal anodal tDCS applied prior to sleep onset reduced total sleep time and significantly increased EEG gamma-band activity, indicating elevated cortical arousal [143]. Notably, this stimulation did not impair overnight memory consolidation [144]. Additional studies have demonstrated that anodal tDCS targeting the left dorsolateral prefrontal cortex (DLPFC) after sleep deprivation reduced subsequent sleep duration across two nights [145]. Similarly, bifrontal tDCS (anode over left DLPFC, cathode over right DLPFC, 2 mA for 30 min) following acute sleep deprivation alleviated subjective sleepiness and fatigue, improved post-deprivation cognitive performance [146], and produced relatively sustained cognitive benefits [146], [147]. These findings suggest that anodal tDCS administered during wakefulness, rather than sleep, may represent a viable strategy for maintaining alertness and enhancing cognition.

In addition, tES has been shown to facilitate sleep onset and strengthen sleep homeostasis. Beyond its role in memory consolidation, slow-oscillatory tDCS (0.75 Hz) administered during SWS reduced the decline of SWO power later in the night [148]. Clinical investigations further indicate that tDCS can improve sleep homeostasis in patients with chronic insomnia, as evidenced by shortened N1 sleep and extended N3 sleep duration [149]. These improvements may also be mechanistically linked to enhanced declarative memory consolidation. Moreover, brief bifrontal repetitive tES (0.75 Hz) applied 30 minutes before bedtime significantly reduced sleep onset latency in individuals with sleep-onset insomnia [150]. Similarly, 5 Hz transcranial alternating current stimulation (tACS) applied over the DLPFC of healthy participants for 15 minutes before sleep accelerated the transition from wakefulness to NREM sleep [27], accompanied by robust entrainment and amplification of EEG theta-band power, which may reflect an enhanced homeostatic sleep drive [151].

#### 2) The feasibility of closed-loop integration

In comparison with other approaches, tES provides a more direct means of modulating sleep through the regulation of cortical excitability. Its safety profile is rigorously supported by real-world clinical evidence-based medicine. For tDCS, stimulation parameters within the ranges of ≤40 minutes, ≤4 mA, and ≤7.2 Coulombs are generally regarded as safe [152], [153]. Nevertheless, tES involves a large number of adjustable parameters, including stimulation target, electrode

TABLE V Transcranial Electrical Stimulation Research Collation.

| Method | Study design | Stimulus parameters | Subject | Result |
|--------|--------------|---------------------|---------|--------|
| tES | Pre–post contrast | intermittently tDCS 30 min after the subject entering SWS, electrodes （8 mm diameter） bilaterally at frontolateral locations and mastoids, positive polarity at both sites (15 s on, 15 s off; current density, 0.26 mA/cm²) | 30 healthy males (mean age 23.8) | ~90 min nap, sleep depth(+), <3 Hz slow oscillatory activity(+), declarative memory(+) [58] |
| tES | Pre–post contrast | 4 min of stable NREM for the first time, start 0.75 Hz sotDCS, five 5-min intervals separated by 1-min intervals free of stimulation, electrodes (8 mm diameter) applied bilaterally at frontolateral locations and mastoids | 13 healthy subjects, 6 male, 7 female (mean age 23.8) | Whole night, SWS(+), slow spindle activity (+), slow oscillations activity(+), declarative memory(+) [141] |
| tES | Within-subject, repeated-measures | before sleep, 2 blocks of 13 min with fade-in/fade-out and 20 min inter-stimulation interval, constant current of 1 mA, bi-frontal target electrodes (5 × 7 cm, FP1/FP2) and bi-parietal return electrodes (10 × 10 cm, P3/P4) | 19 healthy subjects, 13 females, 6 males, (mean age 53.7±6.9) | Whole night, gamma frequency power(+) TSI(-) [143] |
| tES | Placebo-controlled, double-blinded | 30 min tDCS, current 2 mA, density of 0.199 mA/cm², anode electrode placed LDLPFC, cathode placed on the contralateral upper bicep | 36 military subjects, 32 males, 4 females, | 26h sleep deprivation, the second and third night sleep time(-) [145] |
| tES | Randomized controlled, double-blinded | 30 min tDCS, 2 mA, anode on the left DLPFC and cathode on the right DLPFC | 15 healthy males | sleep deprivation, subjective drowsiness and fatigue(-), cognition(+) [146] |
| tES | Placebo-controlled, randomized crossover, double-blinded | NREM 2\3 for at least 8 epochs, start 0.75 Hz sotDCS, five 5-min intervals separated by 1-min intervals free of stimulation, electrodes (8 mm diameter) applied bilaterally at frontolateral locations and mastoids | 6 chronic insomnias Patients, 4 males, 2 females, (mean age 34±7) | Whole night, N1(-), N3(+), sleep efficiency(+) [149] |
| tES | Randomized within-subjects, patient-blind, crossover | 0.75Hz SDR-tES before sleep, for 8 s and then turned off for 10 s, delivered 100 times over a 30-min session, peak amplitudes ranging between 150-500 µA | 12 subthreshold insomnia Patients, 8 females, 4 males, (aged 26-67) | Whole night, SOL(-) [150] |
| tES | Randomized, controlled, cross-over | 15 min, 5 Hz tACS before sleep, electrodes position DLPFC, peak current 1.4 mA | 30 healthy subjects, 12 females, 18 males, (mean age 22.1±3.8) | 15 min nap, theta band total power(+), 1-7 Hz low-frequency power(+), alpha activity(-), awake to NREM earlier [27] |

SWS: Slow-Wave Sleep; TSI: Total Sleep Time Index; SOL: Sleep Onset Latency.

configuration, current density, frequency, and amplitude, which introduces substantial variability and uncertainty in modulation outcomes [154], [155]. As a result, conventional rule-based control strategies frequently converge on suboptimal solutions. Integrating tES with RL for parameter optimization, therefore, constitutes a promising avenue for advancing precision in sleep modulation. Table V summarizes the Transcranial Electrical Stimulation research we have compiled.

### F. Other methods and future direction

Beyond the aforementioned methods, research has also explored the potential of techniques such as transcranial magnetic stimulation (TMS) [156], transcutaneous auricular vagus nerve stimulation (ta-VNS) [157], transcranial ultrasound stimulation (TUS) [158], and transcranial temporal interference stimulation (TIS) [159] for sleep modulation. However, TMS encounters significant challenges for near-term implementation in home environments due to its bulky form factor and high operational costs. The effectiveness of ta-VNS, TUS, and TIS requires further supporting evidence from research. The selection among these methods in practical applications necessitates alignment with specific modulation requirements.

Although recent research has proposed several potentially effective approaches to sleep modulation, the precise therapeutic targets, optimal temporal windows, and corresponding dose-response relationships of each method remain insufficiently characterized, limiting the ability to develop personalized intervention protocols. The inherently low spatial resolution of EEG further constrains accurate source localization of intracranial neural activity, restricting verification of modulation effects primarily to behavioral indices and EEG spectral analyses. In contrast, functional magnetic resonance imaging, with its superior spatial resolution, when combined with EEG or other modalities, enables precise localization of neural targets influenced by stimulation, thereby offering critical insights for optimizing stimulation parameters and timing.

Furthermore, prior studies have predominantly focused on single modulation modalities, neglecting in-depth analysis of potential synergistic or antagonistic effects arising from the combined application of various techniques. Consequently, the development of multimodal sleep closed-loop modulation systems likely represents the primary focus for the next phase of work [160], [161]. Moreover, based on our compiled tabular data, evaluations of modulation efficacy currently lack standardized criteria. We recommend employing PSG for rigorous objective assessment concurrently with validated subjective sleep quality questionnaires. Finally, existing experiments generally lack high-quality evidence-based medical validation. This limitation may stem from multiple factors, including subject heterogeneity, non-standardized experimental designs, and failure to account for the influence of uncontrolled parameters on modulation outcomes. These factors collectively contribute to a strong placebo effect [162], [163] as well as the influence of sleep-related expectations on outcomes [164]. To address these limitations, future work must integrate closed-loop paradigms to enhance methodological rigor and prioritize the implementation of large-scale, multi-center randomized clinical trials in well-defined target populations.

## IV. THE CHALLENGE OF TRANSITIONING TO SLEEP CLOSED-LOOP MODULATION

Section III presents a comprehensive synthesis of modulation strategies suitable for closed-loop paradigms and delineates critical issues that underpin future methodological advancement, thereby providing practical guidance for researchers in selecting and optimizing appropriate approaches. However, critical challenges persist in developing user-friendly systems with high-fidelity monitoring and effective modulation. This work conceptualizes these challenges into three core dimensions: Sensing Solution Selection: Physiological signal acquisition methodologies; Monitoring Model Design: Physiological parameter extraction and state recognition; Modulation Scheme Design: Dynamic optimization of stimulation protocols based on real-time biomarkers. Subsequent sections analyze each challenge and propose potential solutions.

### A. Sensor solution selection

Wearable sleep monitoring systems capture diverse biosignals, including EEG [165], EOG [166], ECG [167], photoplethysmography (PPG) [168], and actigraphy [169]. Each modality serves distinct functional roles: for example, actigraphy is primarily used to distinguish wakefulness from sleep, whereas ECG provides insights into cardiac dynamics during sleep. Variations in signal acquisition techniques and sensor placement further differentiate monitoring strategies. Consequently, two fundamental considerations guide the design of sensing strategies: (i) the selection of biosignals to be monitored and (ii) the methodology for their acquisition. In the context of sleep closed-loop modulation, sensing strategies must meet two essential criteria: high accuracy, reflected in the reliable extraction of sleep-related parameters and waveforms, and low subject burden, achieved by minimizing interference with natural sleep processes. Five-stage sleep classification accuracy is generally regarded as a baseline requirement for modulation optimization (Table VI), and this benchmark serves as the basis for subsequent evaluations.

#### 1) EEG as the core modality for sleep monitoring

Recent progress in EEG-based wearable technologies highlights their strong potential as alternatives to PSG for at-home sleep monitoring. EEG remains an indispensable modality for reliable sleep staging [25], [170], serving as the primary reference signal in clinical PSG [171]. Compared with wearable systems that rely on alternative biosignals, EEG-based platforms consistently achieve superior accuracy in detecting and differentiating sleep stages [172]. This advantage arises from EEG's direct measurement of population-level neuronal dynamics, with each sleep stage characterized by distinct electrophysiological signatures.

#### 2) Advances in prefrontal EEG configurations

Recent survey report that prefrontal electrode configurations are used in nearly 60% of wearable EEG sleep monitoring devices [173], underscoring their increasing adoption in recent system designs [174], [175], [176], [177], [178], [179], [180]. Owing to its anatomical proximity to cortical sources and favorable dermatological features, specifically the smooth, planar skin surfaces that enhance electrode-skin contact, this configuration improves signal-to-noise ratios while simultaneously enhancing user comfort. In addition, prefrontal EEG recordings inherently capture EOG components due to orbital proximity, thereby facilitating discrimination between REM and NREM stages [181]. Multichannel prefrontal configurations further enable comprehensive capture of critical polysomnographic markers, including EOG artifacts, sleep spindles, K-complexes, slow waves, and cortical activation patterns [182]. Importantly, inter-rater agreement for prefrontal EEG staging has been shown to be comparable to manual PSG scoring [178], [183], [184]. Collectively, these findings position prefrontal EEG as the preferred sensing modality for sleep staging in closed-loop modulation, a trend already reflected in its widespread use across existing studies [16], [18], [20], [179], [185]. Moreover, its configuration readily accommodates integration with actigraphy or PPG modules, thereby expanding the diversity of accessible physiological signals.

### B. Monitoring model design

Following biosignal acquisition, monitoring models must extract biomarkers to inform stimulation triggering and optimization. As previously established, heterogeneous biomarkers across systems preclude universal model architectures [186]. This section examines sleep staging models, the most prevalent implementation.

Extensive research has been conducted on sleep stage classification, spanning traditional feature-based machine learning methods to recent deep learning techniques. Nevertheless, accurate and robust sleep staging remains an unsolved challenge, particularly in the context of sleep monitoring and modulation systems. A key reason for this limitation lies in the often formulaic and fragmented nature of existing studies, which lack a systematic framework and clear application focus. Taking sleep closed-loop modulation as an example, we argue that future methodological efforts should be systematically oriented along four critical dimensions: real-time capability, lightweight design, transferability, and interpretability.

#### 1) Real-time capability

Real-time operation is essential for dynamic adjustment of stimulation. While traditional models often rely on recurrent neural networks to exploit sequential dependencies, this architecture introduces substantial latency [187], [188], [189]. A practical alternative is the use of sliding windows that incorporate both historical and current data. Most implementations adopt 30-second epochs as the minimum real-time unit [16], [190], [191], [192], although some researchers argue that this duration exceeds temporal requirements [193]. Ultimately, the optimal epoch length remains application-dependent, determined by the precision required for stimulation timing.

*2) Lightweight design*

Computational and memory constraints of mobile platforms necessitate model minimization. Here, the cloud-based deployment strategy was explicitly excluded from consideration. This decision stems from the impracticality of guaranteeing stable and continuous network connectivity in extreme environments, which significantly limits the applicability of cloud-dependent architectures in such critical scenarios. Moreover, an Edge-based strategy is essential to ensure Real-time response, reliability, data security, and power efficiency in critical monitoring and modulation applications. Preliminary lightweight architectures have been proposed [194], [195], [196], [197], [198], [199], [200], [201], [202], [203]. The authors also proposed a small-scale real-time sleep staging model, namely Micro SleepNet, which has been deployed on mobile terminals [191]. Design priorities vary by platform, with resource-rich devices (e.g., smartphones and tablets) emphasizing performance optimization, while microcontroller units demand strict size constraints.

*3) Transferability*

Transferability represents another critical design priority. Substantial distributional shifts exist between PSG and wearable recordings in closed-loop contexts, owing to differences in electrode placement, signal-to-noise ratios, and participant characteristics [204]. As a result, PSG-trained models often generalize poorly to wearable data, necessitating personalized adaptation. This challenge is further complicated by the scarcity of labeled data in real-world deployments. Consequently, supervised transfer learning approaches, such as fine-tuning, have limited applicability [204]. In contrast, semi-supervised and unsupervised domain adaptation techniques have demonstrated stronger potential under such label-deficient conditions [205], [206], [207], [208], [209].

*4) Interpretability*

Interpretability also remains essential [210]. Attention heatmaps have successfully visualized temporal [191], [211], [212], [213], and time-frequency [214] feature importance aligned with clinical staging criteria. This approach provides more intuitive clinical validation than feature contribution rankings [215], emerging as a preferred explanation strategy.

Finally, recent developments in EEG foundational models offer new directions for sleep staging [216], [217], [218], [219]. Their superior feature representations and transferability may improve performance and facilitate cross-domain generalization. However, further research is needed to evaluate model compression techniques and assess their feasibility for real-time embedded deployment.

*5) Advancing sleep monitoring model by quantitative and fine-grained biomarkers*

At present, the five conventional sleep stages remain the most widely utilized biomarkers in sleep modulation. However, these discrete and non-quantitative categorical labels are inadequate for capturing the full complexity and richness of physiological signals during sleep. As a result, precise modulation of sleep requires the development and application of more sophisticated biomarkers. Future research should prioritize the identification of quantitative, multidimensional, and fine-grained biomarkers, alongside the establishment of robust mappings between these indicators and underlying physiological processes of sleep. Such advancements would deepen the understanding of neural mechanisms during sleep and thereby enhance both the precision and individualization of modulation strategies. Illustrative examples include the sleep depth index [220] for reflecting sleep depth and Delta-Beta phase-amplitude coupling [221] for quantifying sleep quality.

*C. Modulation scheme design*

The modulation strategy constitutes the central element of sleep closed-loop modulation systems, with its accuracy and temporal responsiveness directly determining intervention efficacy. Existing strategies are predominantly grounded in rule-based heuristics and the subjective intuition of system designers [16], [17], [18], [19], [20], [21], [22], [58], [59], [62], [63], [64]. We systematically reviewed prevalent approaches and assessed their conformity to the formal definition of closed-loop operation (Table VI). Findings indicate that most current methods fall within the category of non-adaptive closed-loop strategies, while several studies employed only state-dependent modulation without incorporating feedback-driven parameter optimization, and therefore cannot be considered true closed-loop implementations [17]. At present and for the foreseeable future, a significant barrier remains the inability to precisely physiologically model the effects of intervention methods on human sleep [222]. Consequently, intervention strategies are typically treated as empirically tuned systems lacking explicit physiological interpretability. RL provides a principled framework to address this challenge by enabling data-driven discovery of adaptive policies through iterative interaction with the environment.

*1) RL as a framework*

Drawing from principles of behavioral psychology, RL addresses the interaction between an agent and its environment [223], with its central challenge being the balance between exploration and exploitation. RL has demonstrated remarkable success in achieving superhuman performance across a wide range of games [224], [225], [226], [227] and has more recently been applied within the domain of neural modulation [228], [229], [230], [231], [232], motivating its integration into the design of precise sleep intervention strategies [16] (Fig.6). By defining an appropriate reward function, RL enables continuous exploration of novel parameter configurations while evaluating user feedback in a trial-and-error manner, thereby iteratively optimizing sleep intervention strategies and refining heuristic-based parameter sets that have demonstrated stable effects. For example, Nan Che et al. proposed a lightweight Q-table-based

TABLE VI COMMON MODULATION SCHEMES IN SLEEP CLOSED-LOOP MODULATION.

| Monitoring algorithm | Modulation Strategy | 1 | 2 | 3 |
|---|---|---|---|---|
| Power ratio of theta and alpha bands in EEG | A theta/alpha ratio >1.3 in three consecutive 10-second EEG epochs triggers 10 minutes of tACS and pink noise stimulation, with deactivation within 8 hours after triggering [17] | √ | √ | × |
| Sleep staging, PoAs | The probability of being asleep (PoAs) is calculated in real time, and the audio content recommendation algorithm, instantiated through reinforcement learning, dynamically switches music [16] | √ | √ | √ |
| Sleep staging | Aromatic stimulation is released after detecting stages W, N1, and N2, and stops after entering stages N3 and REM [18] | √ | √ | √ |
| Sleep staging, Power ratio of beta and theta bands in EEG | Modulating the volume of pink noise based on the power ratio of the beta and theta bands in EEG signals [19] | √ | √ | √ |
| Sleep staging | White noise is played to help transition from a wakeful state to light sleep, followed by pink noise to induce deep sleep. Upon waking, white noise is played again to transition from deep sleep to light sleep, while the light strip gradually brightens to help gradually [20] | √ | √ | √ |
| Sleep staging | After falling asleep, the lights turn off and the sleep-aid audio fades away. If wake up during the night, the lights turn on and automatically turn off again after you fall back asleep. When approach preset wake-up time during a period of light sleep, the alarm sounds, the lights turn on [21] | √ | √ | √ |
| Sleep staging | Generate appropriate binaural beat stimulation according to different sleep stages [22] | √ | √ | √ |
| Sleep staging | When SWS reaches 30 seconds, trigger tDCS stimulation and keep it for 30 minutes, then wake up after the first NREM-REM stimulus cycle ends [58] | √ | √ | √ |
| Sleep staging, Spindle wave activity | When NREM stage is detected and spindle activity reaches the defined threshold, apply tACS for 1 second followed by a forced 6.5-second stimulus-free interval. Continue monitoring and execute the above protocol [59] | √ | √ | √ |
| Heart rate | Based on the average heart rate over the previous 5 minutes, subtract 3% to generate vibration stimulation, which is updated every 5 minutes [62] | √ | √ | √ |
| Sleep staging | The frequency of tACS matches the main individualized neural oscillation frequency typical of each sleep period [63] | √ | √ | √ |
| Sleep staging | so-tDCS stimulation is initiated only after entering stage N2 and lasting for 4 minutes, and is applied in blocks. Each stimulation lasts for 5 minutes, with at least 1 minute and 40 seconds of no stimulation between stimulation blocks. Subsequent stimulation blocks are triggered only after entering stage N2/N3 and lasting for 1 minute [64] | √ | √ | √ |

SWS: Slow-Wave Sleep; 1: Real-time monitoring; 2: Modulation affects target state; 3: Feedback Optimization.

RL framework for sleep closed-loop modulation, enabling dynamic optimization of white noise parameters in both temporal and frequency domains [233]. In this section, we first delineate the core RL concepts within the context of sleep closed-loop modulation, followed by a concise overview of its application to the optimization of sleep modulation parameters.

*2) Core components of RL in sleep closed-loop modulation*

Formally, a RL process can be modeled as a Markov Decision Process, defined by the tuple:

$$M = \langle S, A, P, R, y \rangle \qquad (2)$$

Where $S$ denotes the state space, $A$ represents the action space, $P$ defines the state transition probability distribution $P(s_{t+1} \mid s_t, a_t)$, $R$ specifies the reward function $R(s_t, a_t, s_{t+1})$, and $y \in [0,1]$ is the discount factor that determines the trade-off between immediate and long-term rewards.

In RL, the agent is parameterized by policy $\pi(a_t \mid s_t)$, corresponds to the decision-making module that learns and executes the closed-loop intervention policy. Environment encompasses the coupled physiological-physical system, comprising (i) subject-dependent physiological signals (EEG, ECG, respiration) and (ii) external contextual conditions (light, noise, temperature). State ($s_t$) provides a full representation of the system at epoch $t$, though only partial observations $o_t$ (e.g., EEG features contaminated by artifacts) are typically available. Action ($a_t$) encodes the intervention applied at epoch $t$, such as stimulus amplitude, frequency, or duration. External environmental parameters may themselves form part of the action; for example, when light stimulation is employed, its intensity, frequency, and wavelength simultaneously represent environmental parameters and the outcomes of actions applied to preceding states. The set of all feasible interventions constitutes the action space $A$, which in sleep modulation must be carefully constrained to ensure safety, such as by defining maximum permissible intensities or limiting the duration of single stimulation events to prevent harm.

*3) Reward function design and practical implementation considerations*

The reward function $r_t = R(s_t, a_t, s_{t+1})$ quantifies the effectiveness of interventions and serves as the optimization objective in RL. Rewards should be positive (e.g., increased slow-wave sleep, reduced nocturnal awakenings, or improved memory performance). To incorporate broader temporal perspectives, the reward function may integrate long-term outcomes across multiple timescales, including objective measures such as sleep efficiency and the proportion of deep sleep, as well as subjective indicators of daytime functioning. Such metrics enable evaluation of whether multi-night intervention strategies improve the user's overall sleep quality and cognitive or physiological recovery. Here, in Algorithm 1, we present the pseudocode for RL-based sleep closed-loop modulation.

As an illustrative example, a formulation to guide the agent toward physiologically desirable sleep-stage transitions (e.g., progression from light sleep to deep sleep) while maintaining the stability and duration of the target stage once attained. To this end, an reward function can be defined as follows:

$$r_t^{\text{stage}} = \sum_{k=1}^{K} w_k \, P_t(s_t = S_k) - \sum_j k_j \, c_j(a_t) \qquad (3)$$

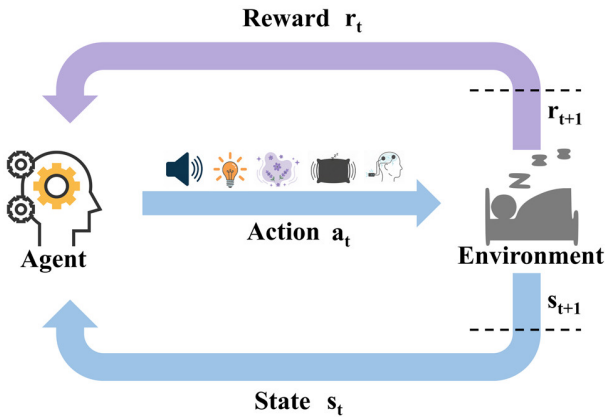

Fig.6. Reinforcement Learning-Based Sleep Modulation Strategy.

---

**Algorithm based on Reinforcement Learning**

1: Initialize state $s_0$ and policy $\pi$
2: t = 0
3: **while** the user is asleep **do**
4:    Observe current state $s_t$
5:    Select action $a_t \sim \pi(s_t)$
6:    Execute action $a_t$
7:    Wait for $\Delta t$ (Update interval $\Delta t$) and acquire new state $s_{t+1}$
8:    Calculate reward $r_t = R(s_t, a_t, s_{t+1})$
9:    Update the policy $\pi$ using the transition tuple ($s_t$, $a_t, r_t, s_{t+1}$)
10:    t = t + 1
11: **end while**

---

Where $S_k \in \{W, N1, N2, N3, REM\}$ denotes the $k$-th sleep stage, and $P_t(s_t = S_k)$ represents the posterior probability of the system being in stage $S_k$ at epoch $t$, as inferred from the sleep-stage classifier or decoding model. The term $w_k$ denotes the weight assigned to each stage, reflecting its physiological desirability (e.g., $w_{N3} > w_{N2} > w_{N1}$). $c_j(a_t)$ denotes penalty terms associated with actions (e.g., excessive stimulation or transitions inducing microarousals), scaled by coefficient $k_j$.

While this formulation provides a principled framework for RL-driven optimization, its direct application in human subjects raises significant ethical considerations and necessitates rigorous safety oversight. A pragmatic approach is therefore to first perform preliminary studies in non-human primates (e.g., macaques) to establish baseline effective parameters, which can subsequently guide the adaptation and validation of interventions in humans. This staged approach ensures both safety and translational relevance.

Furthermore, the rapid advancements in large language models (LLMs) have opened up promising avenues for augmented RL [234]. Leveraging their general world knowledge and robust in-context learning capabilities, LLMs hold potential to be applied in multiple aspects of R L based sleep modulation: aiding in the modeling and selection of sleep modulation strategies, assisting in the design of reward functions, and even simulating modulation processes based on world models. Additionally, LLMs have the potential to integrate diverse process information throughout the sleep modulation process, thereby offering insights into changes in physiological states following the implementation of modulation.

We contend that the application of RL offers promise for addressing current limitations in defining optimal intervention strategies and may significantly advance the field of sleep modulation.

### D. Others

While the preceding discussion has centered on three principal issues, several additional considerations merit attention. Addressing design constraints within intervention protocols represents a key priority for future research. In addition to reinforcement learning, alternative optimization approaches, including Bayesian optimization and evolutionary algorithms, may be applied to refine intervention strategies [235]. When electrical stimulation is utilized as the intervention modality, a major technical challenge arises from the need to suppress stimulation artifacts in simultaneously recorded EEG signals [236], a prerequisite for effective closed-loop implementation [237]. Furthermore, signals collected from wearable devices must be rigorously validated against the gold-standard PSG to ensure reliability [185]. Finally, for overnight monitoring to remain both feasible and comfortable, the selection of user-friendly electrode designs suitable for extended use is of critical importance [238].

## V. CONCLUSION AND OUTLOOK

This review establishes concepts in sleep closed-loop modulation and provides comprehensive methodologies for implementing sleep closed-loop modulation systems. We analyze limitations inherent in traditional open-loop paradigms and formally define sleep closed-loop modulation for the first time. Five classes of methodologies applicable to closed-loop systems are systematically evaluated, and three core implementation challenges are examined with representative case studies. Promising research opportunities, such as developing multimodal modulation systems, are identified. These contributions accelerate the paradigm shift in sleep modulation research.

Future investigations should broaden the scope of closed-loop modulation systems. We propose extending beyond strictly hard real-time implementations toward soft real-time approaches, for instance, by initiating modulation during wakefulness or pre-sleep intervals and subsequently refining protocols using data acquired from overnight monitoring. Continuous daytime tracking of activity patterns and neurobehavioral states via wearable technologies (e.g., wristbands) should also be incorporated to enable predictive modeling of nocturnal sleep dynamics. Such strategies align with the broader objective of developing personalized, sleep-based interventions that support cognitive function, performance, and overall well-being. Notably, recent evidence [239] indicates that targeted neuromodulation may augment cognitive performance and potentially substitute for daytime naps.

We anticipate that sleep closed-loop modulation will play an important role as an adjuvant therapy in sleep medicine while also facilitating the widespread adoption of home-based sleep monitoring and intervention. This trajectory represents a critical step toward proactive, personalized healthcare and may soon render such systems integral to routine sleep health management.

## ACKNOWLEDGMENT

The author would like to extend sincere appreciation to Professor Xuelong Tian of Chongqing University for his invaluable guidance in shaping the author's academic attitude. We used generative artificial intelligence to check grammar and refine the language.